\documentstyle[12pt]{article}

\def\beq{\begin{equation}}
\def\eeq{\end{equation}}
\def\bqn{\begin{equationarray}}
\def\eqn{\end{equationarray}}
\def\bib{\noindent \hskip -0.5cm}
\def\ga{\mathrel{\mathchoice
{\vcenter{\offinterlineskip\halign{\hfil
$\displaystyle##$\hfil\cr>\cr\sim\cr}}}
{\vcenter{\offinterlineskip
\halign{\hfil$\textstyle##$\hfil\cr>\cr\sim\cr}}}
{\vcenter{\offinterlineskip
\halign{\hfil$\scriptstyle##$\hfil\cr>\cr\sim\cr}}}
{\vcenter{\offinterlineskip
\halign{\hfil$\scriptscriptstyle##$\hfil
\cr>\cr\sim\cr}}}}}
\def\la{\mathrel{\mathchoice
{\vcenter{\offinterlineskip\halign{\hfil
$\displaystyle##$\hfil\cr<\cr\sim\cr}}}
{\vcenter{\offinterlineskip
\halign{\hfil$\textstyle##$\hfil\cr<\cr\sim\cr}}}
{\vcenter{\offinterlineskip
\halign{\hfil$\scriptstyle##$\hfil\cr<\cr\sim\cr}}}
{\vcenter{\offinterlineskip
\halign{\hfil$\scriptscriptstyle##$\hfil
\cr<\cr\sim\cr}}}}}

\begin{document}


\begin{flushright}
\noindent{\tt astro-ph/9610092}\\
To appear in {\it The Astrophysical Journal}
\bigskip
\end{flushright}

\begin{center}
{\Large\bf On the Significance of Population II $^6$Li 
Abundances}
\vskip 0.5cm

\noindent 
Martin Lemoine$^{1}$, David N. Schramm$^{1,2,3}$,
James W. Truran$^{1}$, and Craig J. Copi $^{1,2,3}$
\end{center}
\vskip 0.5cm

\noindent
{\it 
1 Department of Astronomy and Astrophysics, Enrico Fermi 
Institute, The University of Chicago, Chicago IL60637--1433\\
2 NASA/Fermilab Astrophysics Center, Fermi National Accelerator 
Laboratory, Batavia, IL 60510-0500 \\
3 Department of Physics, Enrico Fermi Institute, The University 
of Chicago, Chicago, IL 60637-1433}
\vskip 0.7cm

\noindent
{\leftskip 1cm
\rightskip 1cm

\noindent
{\small{\bf Abstract.} 
We study the significance of $^6$Li abundances measured in
metal--poor halo stars. We explore possible depletion factors for
$^6$Li, defined as the ratio of the proto-stellar to the 
observed abundance, in the three stars where it has
been detected; to this end, we assume that $^6$Li/H scales as
$^{16}$O/H throughout the galactic evolution. This assumption
is motivated by the recent observations of a similar
scaling for $^9$Be, and by the fact that, apart from 
$\alpha-\alpha$ fusion creation of $^6$Li, both elements are 
only created in p,$\alpha$--C,N,O spallation reactions.
We examine possible uncertainties attached to the
observations and to the modeling of $^6$Li galactic evolution;
notably, we include a recent evaluation of the primordial production
of $^6$Li.
The depletion factor $D_6$ in the hottest turn-off star HD84937
 is constrained to $D_6\leq4$; this implies that at least one 
star on the Spite plateau has not depleted its primordial $^7$Li 
by more than a factor 4, even by extreme dilution. This constraint 
is in fact stronger when one takes all constraints into account; 
indeed, no current stellar model is able to reproduce the abundances 
of $^7$Li as a function of metallicity and effective temperature, 
and yet allow $D_7\geq2$, while keeping $D_6\leq4$. Therefore,
$^7$Li should not be depleted by more than a factor $\simeq2$
on the Spite plateau. If direct nuclear
burning were the depletion mechanism, then $^7$Li would be depleted 
by less than 2\%. Moreover, all three $^6$Li observations are in
excellent agreement with all standard expectations:  Big-Bang 
nucleosynthesis with $2<\eta_{10}<6.5$, and 
standard stellar isochrones of $^6$Li survival in metal-poor 
stars. We also discuss possible deviations from our assumption 
of a scaling of $^6$Li/H with $^{16}$O/H, due to $\alpha-\alpha$ 
creation of $^6$Li in various sites.}

}
\vskip 1cm

\noindent Subject Headings:
{\it Cosmology: Nucleosynthesis --
Stars: Abundances --
Galaxy: Evolution}
\vskip 1cm

\newpage

\section{Introduction}

 The light nuclide $^6$Li is not significantly produced in
standard Big--Bang nucleosynthesis, ($^6$Li/H)$_p\sim10^{-14}$ 
(Thomas et al., 1993; see below for a new evaluation), and is 
expected to be produced
over the lifetime of the Galaxy in galactic cosmic rays (GCR)
p,$\alpha$--C,N,O spallation, as well as $\alpha-\alpha$ fusion
reactions (Epstein, Arnett \& Schramm, 1974; Reeves, 1994,
for a review). Its high fragility to 
stellar processing makes it a powerful tool to constrain 
Big--Bang nucleosynthesis (Epstein, Arnett \& Schramm, 1974),
as was pointed out by Brown \& Schramm (1988). It was argued 
that the survival fractions $g_6$, $g_7$, of $^6$Li and $^7$Li, 
defined as the ratios of the observed to the proto-stellar
abundances ({\it i.e.} $\equiv1/D$, where $D$ is the depletion 
factor), should be related by $g_6\approx g_7^\beta$, where 
$\beta\sim60-76$ is the ratio of the destruction rate of $^6$Li 
to that of $^7$Li, in the case of nuclear destruction.
Since $g_7=0.99$ already corresponds to a factor $\simeq2$
destruction for $^6$Li, and since the expected proto-stellar
abundance of $^6$Li in metal--poor stars is expected to be at
the limit of feasible detection (see below), the detection of
$^6$Li in a metal--poor halo star could then be turned into a
strong qualitative argument for the absence of significant (nuclear)
depletion of $^7$Li. In that case, the uniformity of the $^7$Li
abundances observed in metal--poor dwarfs on the so--called
Spite plateau (Spite \& Spite, 1982) should indeed reflect the
primordial abundance of $^7$Li. Such an abundance,
($^7$Li/H)$_p\sim1.6\times10^{-10}$, would be 
in excellent agreement with standard Big--Bang nucleosynthesis
and the inferred primordial abundances of D and $^4$He (Copi et 
al., 1995, and references therein). It should be noted that models 
where the surface abundances are depleted by mixing, and thus 
dilution with regions devoid of either $^6$Li of $^7$Li, are limited 
directly by the depletion factors.

It is extremely difficult to detect $^6$Li in population II
(PopII) stars, due to the weak isotopic separation
of the $^6$Li-$^7$Li $\lambda6708$\AA~lines, as compared to 
the width of the $^7$Li line, and to the low 
expected isotopic ratio, $^6$Li/$^7$Li$\la0.1$, notwithstanding
any possible destruction of $^6$Li. A first detection was
nonetheless reported by Smith et al. (1993) in HD84937,
$^6$Li/$^{6+7}$Li=0.05$\pm0.02$. This result was confirmed 
independently by Hobbs \& Thorburn (1994), who reported
$^6$Li/$^7$Li=0.07$\pm0.03$ in the same star, as well as 
$^6$Li/$^7$Li=0.05$\pm0.02$ in HD201891; another marginal 
detection was reported in HD160617 by Nissen (1995), 
$^6$Li/$^{6+7}$Li=0.017$\pm0.012$. 

Smith et al. (1993) were quick to point out the significance 
of their first measurement in HD84937. Using the observed
abundances of boron and beryllium at low metallicities,
and the (Li/B)$_{GCR}$ production ratio expected
from GCR spallation reactions in a PopII environment
(Steigman \& Walker, 1992), these authors showed that
the isotopic ratio $^6$Li/$^7$Li was in excellent 
agreement with: standard Big-Bang nucleosynthesis, {\it i.e.} 
($^7$Li/H)$_p\sim10^{-10}$, no depletion of $^7$Li, 
possibly little destruction of $^6$Li, and no significant 
primordial production of $^6$Li.
Moreover, these authors showed that in the case of rotational
stellar models, which predict large destruction factors for $^7$Li,
$g_7\sim0.1$, and for $^6$Li, $g_6\la0.04$, the measured isotopic 
ratio in HD84937 implies either one of the following: {\it (i)}
a massive production of lithium isotopes in $\alpha-\alpha$
fusion reactions in a very early phase of the Galaxy; {\it (ii)} 
a very high primordial abundance of $^6$Li and $^7$Li, in a 
non-standard Big-Bang nucleosynthesis scenario; or {\it (iii)} 
{\it in-situ} synthesis of $^6$Li in HD84937, for instance in stellar
flares. We examine these different cases in Sec.3.
Steigman et al. (1993) reached very similar conclusions using the 
observed $^9$Be abundance in HD84937: they found agreement with 
standard Big-Bang nucleosynthesis, with no depletion of $^7$Li and 
$^9$Be, and possibly moderate depletion of $^6$Li, $g_6\ga0.2$, as 
predicted by non-rotating standard stellar models. They also argued
that these $^6$Li observations posed a severe challenge to the
so-called Yale rotational stellar models.

A potential problem with these two approaches is that they rely
entirely on the knowledge of the (Li/Be)$_{GCR}$ production ratio at
low metallicities, and it is now accepted that this ratio is in fact 
very uncertain (Fields et al., 1994, 1995). Indeed, the
naive extrapolation at low metallicities of standard galactic
cosmic ray (GCR) spallation is in apparent conflict with the
observed abundances of $^9$Be and B in PopII stars, since it 
predicts a slope $\simeq$2 for the correlations of Log(Be/H) and
Log(B/H) {\it vs} [Fe/H], whereas a slope $\simeq1\pm0.1$ is 
observed ({\it e.g.} Rebolo et al., 1988, Gilmore et al., 1992a,
b, Ryan et al., 1992, Boesgaard \& King, 1993, and 
references, for Be; Duncan et al., 1996a, b and references, for B).
In order to explain these primary behaviors of $^9$Be and B,
different scenarios have been suggested: some involve
modifying the GCR scenario (Duncan et 
al., 1992; Prantzos et al., 1993; Feltzing \& Gustafsson, 1994),
some others modify the chemical evolution in the halo phase
(Ryan et al., 1992; Fields et al., 1993, Tayler, 1995), while some
others invoke different spallation processes (Cass\'e et al.,
1995; Vangioni-Flam et al., 1995). Moreover, 
whereas Steigman \& Walker (1992) have shown that the
(Li/Be)$_{GCR}$ production ratio depends in a crucial way on the 
metallicity of the medium, Fields et al. (1994, 1995) have shown 
that it also depends quite strongly on the physical parameters 
of the GCR flux ({\it e.g.} spectral exponent and path length).
Overall, it appears that this production ratio can be very different
in the above models of early $^9$Be and B evolution, taking
values in the range $\sim10-1000$.

In the present paper, we offer a simple prescription that 
allows us to circumvent the model dependent uncertainties,
hence to derive empirical proto-stellar $^6$Li abundances.
We simply assume that $^6$Li scales as oxygen at all 
metallicities. Such a 
behavior is indeed observed for $^9$Be, and any 
(p,$\alpha$--C,N,O) spallation process which creates $^9$Be should
also produce $^6$Li, in the same way, and vice-versa
(see also below). We justify this prescription further in Sec.2.
We then construct a predicted
evolutionary curve for $^6$Li, and normalize it to the $^6$Li 
meteoritic abundance. It is then straightforward to obtain the 
possible depletion factors for $^6$Li in HD84937, HD201891, and
HD160617 (Sec.2). We also discuss alternatives to this model 
(Sec.3): these can only arise from
$\alpha-\alpha$ fusion, since this is the only channel whereby
$^6$Li and not $^9$Be can be produced.

\section{Standard expectations}
\subsection{Early $^6$Li evolution}

The yield of $^9$Be in a generic spallation process of flux 
$\phi$ may be written:

\begin{equation}
\frac{dy_{{\rm Be}}}{dt}\approx
\langle \phi_i\sigma_{iZ}^{{\rm Be}}
y_Z\rangle + \langle \phi_Z\sigma_{Zi}^{{\rm Be}}y_i\rangle,
\label{yld_Be}
\end{equation}
where $i$ runs over $p,\alpha$, and $Z$ runs over C,N,O; 
$y$ denotes the abundance with respect to hydrogen; the 
average is performed over energy and source and target 
composition; stopping factors due to energy losses and/or escape
are implicitly included as form factors
on the cross-sections. As has been extensively 
discussed in the literature (Vangioni-Flam et al., 1990; Fields 
et al., 1994), the standard GCR yields are 
proportional to the time derivative of the oxygen abundance 
squared $y_O^2$, giving rise to the slope 2, because the GCR 
flux $\phi_i$ is proportional to the supernova rate, the $y_Z$ 
abundances are proportional to the integrated supernova rate,
and the flux $\phi_Z$ is proportional to the product of these 
two factors. As discussed in Fields et al. (1994) and Lemoine et 
al. (1996), one can escape this slope 2 in at least three ways:
{\it (i)} by adjusting the Z dependence of the
flux $\phi_i$ (Fields et al., 1993); {\it (ii)} by 
keeping $y_Z$ constant (Gilmore et al., 1992b; Feltzing \& 
Gustafsson, 1994; Tayler, 1995); and {\it (iii)} by adjusting the
$\phi_Z$ dependence on the supernova rate (Duncan et al., 
1992; Vangioni-Flam et
al., 1995; Cass\'e et al., 1995). The inclusion of Tayler 
(1995) in category {\it (ii)} is not obvious, due to the 
peculiarity of this model; it is discussed in Sec.3. Note that 
Feltzing \& Gustafsson (1994) have shown that spallation in the 
vicinity of supernovae, as advocated by  
Gilmore et al. (1992b), is probably not energetically viable, 
hence we will refer to such models only as an example of 
category {\it (ii)} above. As well, we note that Lambert (1995) 
proposed to explain the halo abundances of $^6$Li, Be and B
by direct production of these elements in the observed stars 
from cosmic rays impinging on the stellar atmospheres. He showed 
that this scenario was not viable for Be and B, and barely 
able to reproduce the $^6$Li abundances.

It is then straightforward to derive the scaling of $^6$Li with 
metallicity, rewriting Eq.(\ref{yld_Be}) for $^6$Li:
\begin{equation}
\frac{dy_{{\rm Li}}}{dt}\approx
\langle \phi_{\alpha}\sigma_{\alpha\alpha}^{{\rm Li}}
y_{\alpha}\rangle +
\langle \phi_i\sigma_{iZ}^{{\rm Li}}
y_Z\rangle+
\langle \phi_Z\sigma_{Zi}^{{\rm Li}}y_i\rangle.
\label{yld_Li}
\end{equation}
In cases {\it (ii)} and {\it (iii)}, the yields of 
$^6$Li scale as the supernova rate. Only in the first scenario 
does the behavior of $^6$Li deviate from that of a primary 
element. Indeed, appropriately adjusting the flux can make the
$^9$Be correlation go from quadratic to linear, in which case the
$^6$Li correlation goes from linear to $\sim$constant through the 
$\alpha-\alpha$ fusion channel. This is 
easily seen by parametrizing the metallicity as a
power-law in time, $y_{Fe}\propto t^n$ (note that $y_{Fe}$ is
the iron abundance \underline{not} in logarithmic units);
tuning the flux so as to obtain a slope 1 for Be and B then
leads to:
log($^6$Li/H)$\propto\log\left({\rm constant}+{\rm [Fe/H]}\right)$.
However, as we discuss in Sec.3, it is not obvious why and how
such a tuning of the flux could take place in the halo phase.

This brief discussion shows that, under very general conditions, 
$^6$Li should behave as a primary element in the halo phase.
However, our prescription further requires that this scaling 
holds equally well up to solar metallicities, {\it i.e.} that 
the transition from the halo to the disk phase proceeds without
any change in the ($^6$Li/O) production ratio. Here, we distinguish
two different cases: {\it (i)} where GCR are responsible for the halo 
evolution of LiBeB, through some modification of their parameters 
or some modifications of the galactic evolution; and {\it (ii)} 
where another process than GCR accounts for the evolution of LiBeB 
in the halo phase. In the first case {\it (i)}, we note that, for
GCR spallation and fusion, at a given (low) metallicity, although 
($^6$Li/O) depends more strongly than ($^9$Be/O) on the very low
energy part $\sim10$ MeV/n of the GCR spectrum, due to the low energy
$\alpha-\alpha$ cross sections, it depends less strongly on the
spectral slope and the path length, that affect 
the higher energy part of the spectrum. We note as well, that, 
within observational accuracy, the ($^9$Be/O) ratio keeps constant
from the halo to the disk phase (see Fig.\ref{Be_ev}). It is 
therefore reasonable to assume that, in case {\it (i)}, the 
($^6$Li/O) production ratio has not changed significantly between 
the halo and the disk phase. In the second case {\it (ii)}, we can 
reasonably assume that the process responsible for the evolution of 
Be and B in the halo phase, proceeds on into the disk phase, and 
that the production ratio ($^6$Li/O) ratio does not change. This is 
indeed what Fig.\ref{Be_ev} suggests. One knows as well that standard 
GCR spallation/fusion creates $^6$Li as a primary element in both 
halo and disk phase. When mixing these two processes of creation of 
$^6$Li, one mixes two primary processes, and one obtains a 
primary process. In this case, therefore, there is no change of 
slope or of production ratio ($^6$Li/O). 
 
Finally, the constancy of the ratio ($^6$Li/O) ratio 
is shown to hold for the models referred to in  
{\it (ii)} and {\it (iii)}, except Tayler (1995);
in this latter scenario, the ($^9$Be/O) ratio in the halo phase 
is arbitrarily tuned to that in the disk phase, but it is not 
possible to make definite predictions for the (Li/O) ratio. We 
discuss this model further in Sec.3. In any case, we allow,
in assigning error bars for our predicted evolutionary curve, for
some possible (slight) mismatch in the ($^6$Li/O) production
ratio between the halo and the disk phase. 

\subsection{$^6$Li as a primary element}
In order to construct the evolutionary curve of $^6$Li, we use a
simple model of chemical evolution. This model is a closed box,
one zone model based on the Tinsley (1980) formalism,
able to reproduce the trends in C, N, O,
[C/Fe]$\sim$[N/Fe]$\approx0$, [O/Fe]$\approx0.5$
in the halo phase (Wheeler et al., 1989), together with the 
age-metallicity relation, and the solar abundances of C, N, O, 
and Fe at solar metallicity. This model incorporates a GCR code 
for production of LiBeB.  We used this model in order to 
take into account the slope changing effect on the $^6$Li--Fe 
relation in the disk phase, due to the introduction of Type Ia 
supernovae, {\it i.e.} in order to follow [O/Fe]. It is 
sufficient for our purpose to use the standard GCR calculations
of spallation/fusion production of $^6$Li, with a flux 
proportional to the supernova rate, as these calculations yield
the desired primary behavior. More complex issues of galactic 
evolution, such as supernovae ejecting metals to form a hot 
intergalactic gas (Mathews \& Schramm, 1993) are irrelevant here.

We note that the upper limit on the primordial abundance of 
$^6$Li has recently been revised upwards by two orders of
magnitude, through new measurements and limits on the cross-section
d($\alpha$,$\gamma$)$^6$Li at low energies (Cecil et al., 1996).
We included this upper limit in the predicted curve,
{\it viz.} ($^6$Li/H)$_p\leq 1.2\times10^{-12}$, 
although it has no effect on our calculations, since the 
metallicities where $^6$Li has been detected are too ``high'' 
for this primordial component to be relevant. Nonetheless, as we 
show in Fig.\ref{Li6_ev}, this could be of relevance for 
metallicities [Fe/H]$\la-3$, depending of course on the
actual primordial abundance of $^6$Li. This will be discussed in 
detail in Nollett et al. (1997).

We thus construct the $^6$Li evolutionary curve, and normalize 
it to the meteoritic abundance of $^6$Li, as shown in 
Fig.\ref{Li6_ev}. Data points for $^6$Li were taken
from Smith et al. (1993), Hobbs \& Thorburn (1994),
Nissen (1995), and Nissen et al. (1995), and are reproduced in Table 
\ref{Li6_data}. We did not consider in these data upper limits
that were previously obtained by Andersen et al. (1984), Maurice 
et al. (1984), Hobbs (1985), and Pilachowski et al. (1989); 
although these observations certainly represented a challenge at 
that time, the upper limits obtained, $^6$Li/$^7$Li$<0.1$ in
all cases, are unfortunately not stringent enough for our
purpose. Data points for $^7$Li and $^9$Be come from a 
compilation of 
Spite \& Spite (1982), Boesgaard \& Tripicco (1986),
Hobbs \& Duncan (1987), Rebolo et al. (1988),
Pilachowski et al. (1993), Gilmore et al. (1992a, 1992b),
Ryan et al. (1992), Boesgaard \& King (1993),
Smith et al. (1993), Hobbs \& Thorburn (1994),
Thorburn (1994), Duncan et al. (1996a, b), 
Molaro et al. (1995), Nissen et al. (1995),
Rebolo et al. (1995), Spite et al. (1996), 
and Anders \& Grevesse (1989).

The $^7$Li
abundances, [Fe/H], and $T_{eff}$ for the $^6$Li stars were taken
from Smith et al. (1993), Hobbs \& Thorburn (1994), Thorburn 
(1994), Nissen (1995), Nissen et al. (1995), Primas (1995),
Spite et al. (1996), and Ryan et al. (1996); see Table 
\ref{Li6_data}. We adopted, as error on the temperatures, 150K 
statistical, and 100K systematic, at 95\% c.l., following Spite et 
al. (1996). Note that, 
contrary to elemental abundances, the isotopic ratio is only
weakly sensitive to the [Fe/H] and $T_{eff}$ adopted; for
instance, the $^7$Li abundance of HD84937 as derived by Thorburn 
(1994) is 0.24dex higher than that derived by Smith et al. 
(1993), although the isotopic ratios are entirely consistent. 
For this star, both $^6$Li/$^{6+7}$Li evaluations were combined
in a weighted least squares procedure.

We adopted a $\pm0.16$dex $2\sigma$ statistical gaussian uncertainty
for lithium and iron abundances, which is a typical estimate
(Thorburn, 1994; Spite et al., 1996).
We adopted a $\pm0.2$dex 95\% c.l. systematic uncertainty for
lithium and iron abundances, a typical estimate as well 
(Thorburn, 1994; Spite, 1995; Spite et al., 1996).
In Fig.\ref{Li6_ev}, we combined these systematics 
on the $^7$Li and $^6$Li abundances quadratically 
with the 2$\sigma$ statistical error bars, in order to produce 
an estimate of the total 95\% c.l. on these data points. In the 
following analysis, however, we allow for different statistics. 
Namely, we take the abundance systematics on N[Li] and [Fe/H] to 
follow top--hat probablity laws, and the $^6$Li/$^{6+7}$Li ratios 
to follow gaussian laws. 
We adopted a $\pm0.3$dex 95\% c.l. uncertainty on our
evolutionary curve, which we can attribute to the
normalization to the meteoritic abundance of $^6$Li as well as
to slope changing effects of chemical evolution at the halo-disk
transition. Indeed, although the meteoritic abundance
of $^6$Li, as given by Anders \& Grevesse (1989), is endowed
with very small error bars, $\pm0.04$dex, we prefer to remain 
conservative and allow for a possible 2$\sigma$ uncertainty
$\simeq\pm0.2$dex in promoting this meteoritic abundance to a
`cosmic' abundance at [Fe/H]=0. 
We show, in Fig.\ref{Be_ev}, how well this evolutionary curve,
and its accompanying error bars, bracket the Be observations
in halo stars. We always refer to $^9$Be
as a baseline, and not to B, since the available boron
abundances have not yet been fully
corrected for NLTE effects. Let us mention, however, that a
clear primary correlation for boron {\it vs} iron in the halo
phase, with a slope $\simeq1.0\pm0.05$, is inferred from these
latter abundances (Duncan et al., 1996a,b).

We now discuss the three detections of $^6$Li in HD84937, 
HD201891, and HD160617. The upper limits located above the
evolutionary curve simply represent non-detection of $^6$Li,
while those below the curve are interpreted as non-detections
due to destruction of 
$^6$Li in these stars. We will show in later discussion that this is 
compatible with their $T_{eff}$. It is straightforward to 
evaluate the survival fraction of $^6$Li in the three 
stars through statistical bootstrapping, using the estimator:
\begin{equation}
{\hat g}_6\equiv\frac{(^6Li/H)_{obs}}{(^6Li/H)_{pred}}.
\end{equation}

In using this estimator, it is implicitly assumed that no 
production of $^6$Li took place in the star, {\it i.e.} we 
neglect any flare production (Deliyannis \& Malaney, 1995); we 
will return to this point in Sec.3.
In this bootstrapping, we draw at random, according to their 
statistics, the `observed' and the `predicted' abundances.
We draw the predicted proto-stellar abundances for the three stars 
from the same evolutionary curve, so as to ensure that the slope 
of the $^6$Li correlation
{\it vs} [Fe/H] is 1 in the halo phase. We do not  
correlate the systematics of the observed abundances, since 
these stars have different metallicities and different $T_{eff}$.
We took care to include the metallicity statistics of the stars
in the bootstrapping, taking into account the correlation of
($^6$Li/H)$_{pred}$ with metallicity.
The resulting $\hat g_6$ histogram is shown in Fig.\ref{his_g6} 
for a sample size of 10$^6$.

We thus obtain the following averages and 95\% c.l. limits on
$g_6$, in the standard case:\medskip

\begin{center}
\begin{tabular}{lcc}
HD84937 & ${\overline g_6}=1.92$ & $0.57\leq g_6\leq7.24$ \\
HD201891 & ${\overline g_6}=0.14$ & $0.02\leq g_6\leq0.54$ \\
HD160617 & ${\overline g_6}=0.38$ & $0.00\leq g_6\leq1.67.$ 
\end{tabular}
\end{center}\medskip

The lower limit on $g_6$ for HD160617 is 0.00: it could have 
been expected on the basis that $^6$Li was not detected at the 
$2\sigma$ level in this star. We checked that the figures above
were not sensitive to the type of statistics adopted for the
systematics and the theoretical modeling, within $\simeq$10\%.

With our present assumption that flare production of 
lithium did not occur in these stars, ${\overline g}_6>1$ is
unphysical. One usual remedy is to use Bayesian statistical
inference, and to truncate the ${\hat g}_6$ probability with
(for instance) a top hat function 
$\Theta\left({\hat g}_6\right)\Theta\left(1-{\hat g}_6\right)$,
to enforce $0\leq {\hat g}_6\leq1$, and re-normalize the 
${\hat g}_6$
distributions; in that case, the resulting 95\% c.l. limits
would become (0.26,0.99) for HD84937,  (0.03,0.48) for HD201891,
and (0.03,0.91) for HD160617, and the averages: 0.71 for HD84937,
0.14 for HD201891, 0.30 for HD160617. We note, however, that
these numbers are subject to the (subjective) choice of
constraints applied on ${\hat g}_6$, as is the case in any
Bayesian inference procedure. The use of a top-hat uniform law 
is justified as a zero-th order choice when one does not have
any more reliable guess to offer. This is the case here, since
so-called 'standard' isochrones do not agree with each other as
to the degree of destruction of $^6$Li in these stars
(see Chaboyer, 1994, on the one hand, and Deliyannis et al.,
1990, on the other hand).

Another remedy is to invoke potential uncontrolled systematics,
which, in
the present case, should be associated with the estimates of the
$^6$Li/$^{6+7}$Li ratio. Indeed, as pointed out by Nissen (1995),
convective motions in the stellar atmosphere result in 
an asymmetry of the profile, that might mimic $^6$Li absorption 
in the Li{\sc i} line. The choice of the type of absorption 
profile also plays an important role in the estimate of the 
$^6$Li/$^{6+7}$Li ratio. Although these effects were taken into 
account by Nissen (1995), Nissen et al. (1995), using the nearby
Fe{\sc i} and Ca{\sc i} lines as standards, the choice
of these lines is dictated by convenience alone, and some
mimicking of $^6$Li is still possible. Such systematics could be 
ascertained using a greater sample of lines as standards;
obviously, higher spectral resolution at high
signal-to-noise ratio would allow a substantial increase 
in the accuracy of these estimates.
A zero-th order estimate of such systematics is given by the
difference of the observed to the predicted abundance of $^6$Li 
in HD84937, yielding a reasonable offset of $+0.3$dex.
This star is indeed the hottest of the three stars, 
and, for $T_{eff}\simeq6300$, possibly no depletion of $^6$Li 
is expected in stellar models (Brown \& Schramm, 1988;
Deliyannis et al., 
1990, Chaboyer, 1994), depending on the (unknown) evolutionary
status of HD84937, whose photometry puts it at the turn-off.
Note that such systematics would
not be in contradiction with the other two stars, that are
main-sequence dwarfs with $T_{eff}\simeq5800$, for which
moderate to strong $^6$Li depletion is expected. Thus,
if we shift, by hand, the observed data point, to place it
right on top of the predicted curve, the 95\% c.l. limits on 
${\hat g}_6$ in HD84937 become (0.30,3.16).  It is 
comforting to note that both remedies used above, to enforce
${\overline g_6}\leq1$, lead to a very similar 95\% c.l. lower limit
on ${\hat g}_6$.

Finally, in order to check the consistency of the derived 
depletion factors with stellar isochrones for $^6$Li survival, 
we plot in Fig.\ref{Li6_Teff} the observed data points and upper 
limits {\it vs} $T_{eff}$. Unlike $^7$Li, however, for which the 
measured abundances are at the same level on the Spite plateau,
the $^6$Li data points need to be corrected for their trend with
metallicity; in Fig.\ref{Li6_Teff}, they were brought back to
the metallicity of HD84937. It is obvious from 
Fig.\ref{Li6_Teff} that the locus of these data points is
consistent with the expected shape of isochrones, as shown in
Chaboyer (1994) for instance, and, more specifically, that none
of the upper limits derived contradict the three detections. 
Moreover, we did not reproduce the isochrones of Chaboyer on 
Fig.\ref{Li6_Teff}, since most of these stars are located near 
the turn-off, and the isochrone seems to depend strongly on the 
evolutionary status of the star (Chaboyer, 1994).
As well, note that to change the opacities used in the code is 
sufficient to modify the $^6$Li depletion factor from 0 to 
$\simeq3$ (in the subgiant case), which means, conversely, that 
we cannot rely too much on these models. In any case, the 
general agreement is very satisfying.

To conclude this section, we obtain a 95\%c.l. lower limit on
the survival fraction in HD84937, $g_6\geq0.26$ (equivalently
$D_6\leq4$ for the depletion factor), assuming some 
uncontrolled systematics in assessing the line profile of the 
$^{6,7}$Li $\lambda6708$\AA~line, or using Bayesian inference to 
bring back the measured average ${\overline g}_6$ in the physical
region ${\overline g}_6\leq1$. It is well known that $^6$Li is 
much more fragile to nuclear burning than $^7$Li ({\it e.g.} 
Brown \& Schramm, 1988, and references therein). As a consequence,
the preservation region of $^6$Li is, at any time, much shallower
than that of $^7$Li. As a consequence, $^6$Li is also more 
sensitive than $^7$Li to depletion mechanisms such as dilution 
processes, or even mass-loss (Vauclair \& Charbonnel, 1995).
Therefore, it seems reliable to use the lower limit on $g_6$ as an 
extreme lower limit on the $^7$Li survival 
fraction $g_7$. Our evaluations thus show that at least one star 
on the Spite plateau has depleted its $^7$Li by less than a 
factor 4, at 95\% c.l.. Using the observed abundance of $^7$Li 
in HD84937, Log($^7$Li/H)=$-9.77\pm0.16\pm0.2$ (95\% c.l. estimates for 
both statistical and systematics),
we derive an upper limit on the primordial abundances of $^7$Li, 
Log($^7$Li/H)$_p\leq-8.81$ (95\% c.l.), where we added
linearly the errors and the upper limit on the depletion
factor. Although this upper limit is not too stringent in 
itself, we note that, in order to saturate the upper bound, one 
would have to deplete \underline{both} $^6$Li \underline{and}
$^7$Li by the same amount 
(a factor 4), and yet reproduce all the observed features of the 
Spite plateau, namely the uniformity in abundances and the 
isochrones. The above constraint is therefore stronger, {\it i.e.}
it amounts to $D_7\la2$, since there is, at our knowledge, no 
stellar model able to reproduce the observed trends, while having
$D_7\geq2$ and $D_6\leq4$. We note, for instance, that, if the
maximal depletion 
of $^6$Li were due to pure nuclear burning without dilution, then 
$^7$Li would be depleted by less than 2\% from its primordial value
(Brown \& Schramm, 1988). As well, the rotational mixing models 
(Pinseonneault et al., 1992) satisfy the latter constraint on 
the features of the Spite plateau with a primordial abundance
Log($^7$Li/H)$_p\sim-9$, but cannot do without a depletion factor
$D_6\ga 25$ (Deliyannis \& Malaney, 1995).

\section{Alternative scenarios}

As mentioned earlier, the only way $^6$Li can avoid  a 
primary behavior for the whole range of metallicities 
$-4\leq$[Fe/H]$\leq0$, comes through
the $\alpha-\alpha$ fusion channel. To discuss these 
'non-standard' cases, we are compelled to restrain ourselves to 
specific models: arbitrary tuning of the cosmic ray 
flux, referred to as {\it (i)} in Sec.1.1, spallation in 
globular clusters (Tayler, 1995), or stellar flare production of 
$^6$Li (Deliyannis \& Malaney, 1995).

\subsection{Early bright galactic phase}
As discussed earlier, different models have been proposed to tune
the cosmic ray flux so as to 
reproduce the observed abundances of $^9$Be and B in the 
galactic halo. Ryan et al. (1992) suggested an outflow model 
with an enhanced supernova rate in the halo phase; since the 
injection flux is traditionally assumed to be proportional to 
the supernova rate, it was believed that it would allow the 
production of higher Be and B abundances. This is incorrect 
in that increasing the supernova rate leads to an increase in 
the production of metals, so that a higher $^9$Be or B abundance 
would be associated with a higher metal abundance, and no 
difference with the standard model would be noted in an
elemental graph Log($^9$Be/H) {\it vs} [Fe/H]. The higher
$^9$Be and B 
abundances were rather obtained in an {\it ad-hoc} fashion by 
normalizing the yields to the observed $^9$Be abundance at 
[Fe/H]=$-1$; in particular, a slope 2 is obtained. The role of 
the outflow was to evacuate the gas to prevent a high supernova 
rate in the disk phase, so as to prevent an overproduction of 
$^9$Be and B at solar metallicities (see Fig.3 in Prantzos et 
al., 1993 for an illustration of these effects). Therefore, 
since further observations of Pop II $^9$Be and B abundances seem
to confirm a slope of 1 on the whole range of metallicities
$-4\leq$[Fe/H]$\leq$0, we need not consider this model further.
This remark also applies to the models of Prantzos et al. 
(1993), where the path length was tuned in the halo phase so as 
to reproduce the $^9$Be and B abundances; although this model 
was nicely motivated, namely the cosmic rays
should be overconfined in the halo phase, it led to a slope 
$\simeq1.75$ for $^9$Be and B, and, in any case, to a slope 
$\simeq1.0$ for Li.

Finally, in Fields et al. (1993) it was shown that to reproduce 
a slope of 1 for $^9$Be and B, the cosmic ray flux should 
vary approximately inversely with time. These authors did 
not offer an explanation for such a behavior: they were rather 
interested in constraining the duration and luminosity of an 
early bright galactic phase under the assumption that the cosmic
ray flux would effectively vary as 1/$t$. It was shown that the 
bulk of the $\gamma$-ray background could be associated with 
such an enhanced cosmic ray flux in the first few billion years. 
Nevertheless, according to the above discussion, Ryan et al.
(1992) have shown that an increased supernova rate would not
lead to a slope 1, and Prantzos et al. (1993) have shown that an
overconfinement would not do so either. Hence, the only remaining
explanation for a GCR flux varying as the inverse power of time
would be to have the injection flux evolve as a high power of 
the supernova rate, which does not have a physical basis at the
present time. Indeed, with (O/H)$\propto t^n$, $n\la1$,
one would require 
$\phi_{inj}\propto\left(dN_{SN}/dt\right)^{1/1-n}$.
Finally, we note that Lemoine et al. (1996) have shown that the GCR 
propagated flux should evolve as a very low power $<1$ of the 
supernova rate, throughout the disk phase, in order not to
overproduce the LiBeB abundances at solar metallicity. Thus, the 
higher power could only apply in the early pre--disk phase.

\subsection{Spallation in globular clusters}

Tayler (1995) has offered an original explanation to the primary 
behaviors of $^9$Be and B in the halo phase, in suggesting that 
the bulk spallation would take place in globular cluster 
sized objects before the gas would be swept out of the cluster 
and mixed with the ISM gas. The primary behavior arises from the 
assumption that every globular cluster ejects the same amount of 
gas, and that the spallation is the same in every cluster: each 
time metals (Z) are ejected to enrich the ISM, LiBeB elements are
also ejected, with a constant (LiBeB/Z)$_{ejec}$ ratio, that is 
independent of the Z content of the ISM. This ratio should not 
in fact be constant, as these globular clusters should not eject 
exactly the same amount of metals, and the spallation is 
not  expected to be exactly the same in every object, so that,
in the end, one would expect to observe a linear relation
Log($^9$Be/H) {\it vs} [Fe/H] with some scatter. There has indeed 
been a recent claim for a scatter around the linear trend (Primas, 
1995), which gives credit to this scenario.

This model is nonetheless subject to relatively strong requirements.
It assumes that all globular clusters eject nearly 
the same amount of metals, which, in a first approximation, 
would mean that they all have very similar masses. Moreover, it
is assumed that the spallation is much more efficient in the 
globular clusters than in the ISM. Finally, the ($^9$Be/O) 
production ratio has to be arbitrarily tuned to that observed 
in the disk phase, since  the linear 
behavior of $^9$Be {\it vs} [O/H] is observed on the whole range
$-3\la$[O/H]$\la$0 (Fig.\ref{Be_ev}). On this basis, it is
unfortunately
difficult to say whether or not the (Li/O) production ratio is
the same in these globular clusters and in the disk phase,
since the (Li/Be) ratio depends very strongly on the spectral
slope and the confinement time of the accelerated projectiles
(Fields et al., 1993, 1995). Hence, no reliable evolutionary curve 
for $^6$Li can be constructed in this case.

\subsection{Flare production}

Deliyannis \& Malaney (1995) have studied the possibility of 
generating the observed $^6$Li of HD84937 {\it via} flare 
production. They argue that the energetics can, in principle, be 
fulfilled. However, we disagree with their estimate of the 
energetics required, and we show below that such a model is strongly 
disfavored for producing all the observed $^6$Li. These authors also 
propose two observational tests to possibly discriminate {\it in 
situ} flare produced $^6$Li from proto-stellar $^6$Li: {\it (i)} the 
$^6$Li/$^9$Be and B/Be ratios in HD84937; and {\it (ii)} the curve 
of $^7$Li/$^6$Li ratios {\it} the effective temperature in a sample 
of metal-deficient dwarfs. We disagree with these authors on the 
feasibility and the reliability of these tests, as we discuss now.

The first  test of Deliyannis \& Malaney (1995) to distinguish 
{\it in situ} flare produced $^6$Li from proto-stellar $^6$Li relies 
on the difference between the $^6$Li/$^9$Be and B/Be ratios resulting 
from flare production and those resulting from cosmic ray spallation.
However, it is straightforward to show that, even if all $^6$Li 
observed in HD84937 were produced by flares, the amount of Be and B 
produced by these flares would be negligible with respect to the 
amount of Be and B observed in this star. Using the flare production
ratios of Walker 
et al. (1985), $^6$Li/$^9$Be$\sim40-100$ for a solar metallicity star, 
and rescaling it to a metallicity [Fe/H]=$-2.1$, this ratio should 
become $^6$Li/$^9$Be$\sim10^3-10^4$ in HD84937, since $^6$Li is still 
produced through $\alpha-\alpha$ fusion, but the p--C,N,O channel is 
cut off. This ratio has to be compared with the observed ratio 
$^6$Li/$^9$Be$\simeq66^{+90}_{-45}$ (95\% c.l.): clearly, if all 
observed $^6$Li has been produced {\it in situ} by flares, then only a 
tiny (and negligible) fraction of the observed $^9$Be comes from these 
flares. This conclusions holds true for B as well, and therefore, the 
$^6$Li/Be and B/Be ratios cannot be used to trace {\it in situ} 
flare production.

The Deliyannis \& Malaney second observational test rests on the 
dependence of the observed $^7$Li/$^6$Li ratio with the effective 
temperature in subgiants. Deliyannis \& Malaney (1995) have shown that 
this ratio should remain constant if the observed $^6$Li is of 
proto-stellar origin, in the range of temperatures 6400$\geq 
T_{eff}\geq5800$, as neither of both isotopes is then depleted. If
$^6$Li has been created in flares, the isotopic ratio should decrease
with decreasing temperature, in the same range of temperatures, due
to increased dilution ({\it i.e.} larger convection zone) past the 
turn-off. Indeed, the essential difference between proto-stellar and 
flare produced $^6$Li is that the proto-stellar $^6$Li preservation 
region is considerably larger than the convection zone post turn-off, 
whereas the only region containing flare produced $^6$Li is the 
convection zone itself. This behavior is, however, very dependent 
on the evolution of the convection zone, which is difficult to model 
during the ascent on the subgiant branch. Moreover, as pointed out 
before, it seems that the survival of $^6$Li in these stars depends 
strongly on the opacities adopted. Overall, it seems that, even if a 
large database of $^7$Li/$^6$Li ratios in metal-poor stars could be 
collected, a possibility that is already very optimistic, this would 
not readily allow one to discriminate between flare produced and 
proto-stellar $^6$Li.

We now turn to the energetics required for producing $^6$Li in flares 
in HD84937. Deliyannis \& 
Malaney (1995) argue that the production efficiency is 
10$^{-3}$--1 atom/erg, so that 10 flares per year of strength
10$^{32}$ ergs, which corresponds to the large solar proton
flares, for 1 atom/erg, during 1Gyr, would reproduce the
observed $^6$Li. The point is that large proton flares, in the
Sun, are associated with steep power law spectra (Ramaty et al.,
1995), and the efficiency for these spectra is $\simeq0.007$ atom/erg
(with a low-energy cut-off imposed at 1 MeV) for a solar composition
(R. Ramaty, 1996, private communication). We calculate that for a
composition rescaled to the metallicity of HD84937, this estimate
goes down to $\simeq0.003$ atom/erg, assuming a 1 MeV cutoff.
The Deliyannis \& Malaney estimate of 1 atom/erg is therefore too 
optimistic by a factor $\simeq300$, and instead of 10 flares per 
year, $\simeq3000$ flares per year of a strength comparable to 
that of the large solar flares would be required, an enormous 
number indeed.

Here we investigate in greater detail the production of lithium
in flares. We use only the $\alpha-\alpha$ fusion channel in our 
calculations, since with its threshold and peak resonances 
around $\simeq10$ MeV/n, and the low metal content of the star, 
it should largely dominate over the p,$\alpha$--C,N,O channel.
We assume an abundance $n_{\alpha}/n_p=0.1$ in the accelerated 
and target particle abundances, although for large solar flares,
the accelerated abundance could be lower by a factor 10 (Murphy 
et al., 1991). We also assume that, through downward 
convection, all the $^6$Li produced in flares would be mixed 
in the convection zone of mass $M_c$. In a thick target model,
the rate of $^6$Li production can be written (Ramaty \& Murphy, 
1987):

\begin{equation}
Q\left(^6{\rm Li}\right)=
\frac{1}{m_p}\frac{n_\alpha}{n_p}\int_0^{+\infty}
\sigma_{\alpha\alpha}^{^6{\rm Li}}\left(E\right)\left(\frac
{{\rm d}E}{{\rm d}x}\right)_{\alpha}^{-1}
\int_{E}^{+\infty}N_{\alpha}
\left(E'\right)\,{\rm d}E'{\rm d}E,
\label{Q6}
\end{equation}
where $Q\left(^6{\rm Li}\right)$ is in atom.s$^{-1}$, 
$\left({\rm d}E/{\rm d}x\right)_{\alpha}$ represents the energy
losses of $\alpha$ nuclei in the interaction medium, in
MeV.g$^{-1}$.cm$^2$, and $N_{\alpha}\left(E\right)$ is the
injection spectrum of $\alpha$ particles in
(MeV/n)$^{-1}$.s$^{-1}$. This latter can be 
parametrized by a power law, a decreasing exponential in 
rigidity, or a Bessel function K$_2$ of the second kind, 
although we note that this latter reproduces more adequately the 
observational data on solar flares. It is also customary to 
define the deposited power:

\begin{equation}
\dot{W}=\sum_{i}A_i\int_{0}^{+\infty}E\,N_i\left(E\right)\,dE,
\end{equation}
where $A_i$ denotes the atomic mass of nucleide $i$.
The production efficiency, referred to above, is then defined as 
$Q\left(^6{\rm Li}\right)/\dot{W}$, in atom.erg$^{-1}$. Here, 
however, we calculate the rate of production of $^6$Li nuclei, 
using Eq.\ref{Q6}, normalizing the injection spectrum to the 
observed rate of average irradiation of the solar atmosphere, 
{\it i.e.} $N_p\left(>30{\rm MeV}\right)\simeq3.10^{26}$ 
protons.s$^{-1}$ (Ramaty \& Simnett, 1991). We find that the 
production rate does not depend strongly (within a factor 2)
on the type of spectrum chosen, or on the low-energy cut-off
adopted for power law spectra. For power law spectra, we adopt 
a steep spectrum (index $\simeq4$), with a low-energy cut-off of
1 MeV/n. Apart from preventing divergence, this cut-off is to
simulate the presence of energy losses that flatten the spectrum
at low energies.
Due to the normalization adopted above, and the fact that
the threshold of the $\alpha+\alpha\to$ $^6${\rm Li} cross-section
is $\sim10$ MeV/n, our choice of cut-off maximizes the production rate.
The production efficiency is  strongly dependent on the type 
of spectrum, varying between $\sim10^{-3}$ and $\sim10^{-1}$
atom.erg$^{-1}$, in agreement with previous calculations (Canal 
et al., 1975, 1980). 
The production rate obtained is:
$Q\left(^6{\rm Li}\right)\sim2-4\times10^{22}\;
{\rm atom.s}^{-1}$,
which, after mixing in the convective zone, on a production
timescale $\tau=1$ Gyr, yields:
$$\left(\frac{^6{\rm Li}}{{\rm H}}\right)\sim 2-4\times10^{-15}
\left(\frac{M_c}{10^{-3}M_{\odot}}\right)^{-1}
\left(\frac{\tau}{1{\rm Gyr}}\right).
$$
As compared to the observed $^6$Li abundance in HD84937, 
($^6$Li/H)$\simeq 9.\times10^{-12}$, this estimate is negligible.
It has to be noted, however, that the mass $M_c$ of the
convection zone is largely unknown, mostly because the
evolutionary status of HD84937 is itself unknown.
Deliyannis \& Malaney (1995) point 
out that $M_c$ should range from $3\times10^{-4}M_{\odot}$ to
$3\times10^{-3}M_{\odot}$, with an estimated $10^{-3}M_{\odot}$,
if HD84937 is a dwarf, and from $7\times10^{-5}M_{\odot}$ to
$1\times10^{-3}M_{\odot}$, with an estimated $3\times10^{-4}$
if HD84937 is a subgiant. The mass of the convection zone also 
depends strongly on the stellar parameters, such as the mixing 
length, and, overall, the above estimates should be good 
within an order of magnitude. The estimate of a production 
timescale $\tau\approx 1$ Gyr stems from the dependence of the 
convection zone on age: since the convection deepens with age, 
on the main sequence, any spallated $^6$Li is more likely to 
have been produced recently; Deliyannis \& Malaney (1995) 
estimate 1 Gyr to be a reasonable timescale, and we adhere to
this value. 
This shows that, even in the most favorable case (for flare 
production), {\it i.e.} a convection zone
$M_c\sim10^{-5}M_{\odot}$, the $^6$Li abundance  
 produced in flares remains more than an order of magnitude 
below the observed abundance. Recall as well that we assumed 
that the downward convection of flare produced $^6$Li was fully 
efficient, that we assumed a ``maximal'' $n_\alpha/n_p$ 
abundance, and that we maximized the production rate in our choice
of spectrum.

Hence, we feel that flare production of $^6$Li cannot produce 
a significant contribution to the observed $^6$Li abundance, 
unless HD84937 is subject to tremendously powerful flares over 
more than 1 Gyr.

\section{Conclusion}

We have estimated the depletion factor of $^6$Li in the three 
metal-poor halo stars where it has been detected. In order to 
determine the proto-stellar $^6$Li abundance in these stars, we 
have constructed an ``empirical'' evolutionary curve for $^6$Li,
deduced from the observed abundances of $^9$Be. Namely,
we have assumed that $^6$Li/H scales as $^{16}$O/H over the galactic 
lifetime, and normalized the resulting 
evolutionary curve to the meteoritic abundance of $^6$Li. Our 
assumption is motivated by the recent observations of metal-poor 
stars, that revealed a slope $\simeq1$ for the correlation of 
Log($^9$Be/H) with Log($^{16}$O/H). One may indeed recall that
$^6$Li and $^9$Be share their origin in p,$\alpha$--C,N,O
spallation reactions (apart from an extra $\alpha-\alpha$ fusion
channel for $^6$Li).

We have taken care in assigning uncertainties to the so-modeled
evolutionary curve of $^6$Li. Notably, we have included new 
cross-section data for the primordial production of $^6$Li. We form
the ratio of the observed to the proto-stellar abundance to obtain
the survival fraction $g_6$, {\it i.e.} the inverse depletion 
factor; the statistics of this survival
fraction, in each star, are obtained from bootstrapping. We thus
obtain a 95\%c.l. limit $g_6\geq0.26$ (equivalently $D_6\leq4.$ 
for the depletion factor) in 
the hottest turn-off HD84937 ($T_{eff}\simeq6300$K). We 
have argued that some systematics, arising from the inaccuracy 
of stellar line profile modeling (at the precision required), 
should be associated with these measurements of the 
$^6$Li/$^{6+7}$Li ratio, as there is always the possibility that 
some weak profile distorsion might mimic the $^6$Li absorption.
The above upper limit has been obtained in two different ways:
assuming such systematics on the $^{6,7}$Li line profile, and 
proceeding to Bayesian inference, to return the average
${\overline g}_6\simeq2.$ (measured for HD84937) into the physical
region ${\overline g}_6\leq1$. 

Since $^6$Li is much more fragile 
than $^7$Li to nuclear burning, and since the preservation 
region of $^6$Li is always shallower than that of $^7$Li, this 
upper limit on the depletion factor $D_6$ is also an extreme upper
limit on $D_7$, the depletion factor for $^7$Li. At a minimum, this
implies that at least one star on the Spite plateau, with abundance
Log($^7$Li/H)=$-9.77\pm0.16\pm0.2$ (95\% c.l. for each error bars, 
statistical and systematics), has depleted its $^7$Li 
by no more than a factor 4. This, in turn, means that an 
extreme upper bound can be put on the primordial abundance of $^7$Li,
Log($^7$Li/H)$_p\leq -8.81$ (95\% c.l.). We note that,
in order to saturate this upper bound, $^7$Li \underline{and}
$^6$Li would have to have been depleted at the same level, by a 
factor 4. For instance, if depletion had occurred through direct 
nuclear burning, $^7$Li would have to be depleted by no more than 
2\% for $^6$Li to be depleted by a factor 4 (Brown \& Schramm, 1988).
As well, rotational mixing models (Pinseonneault 
et al., 1992), that claim to 
obtain a much higher depletion factor for $^7$Li than `standard' 
models, and yet reproduce all the observed features of the Spite 
plateau, are ruled out: in these models, a primordial abundance 
Log($^7$Li/H)$_p\simeq-9$ is associated with a depletion factor 
$D_6\sim25-40$ (Deliyannis \& Malaney, 1995). We know of no current
stellar model able to reproduce the observed features of the Spite 
plateau, and yet allow $D_7>2$, while keeping $D_6\leq4$.
The above constraint on the depletion factor of $^7$Li on the Spite 
plateau is therefore stronger, {\it i.e.} $^7$Li should not be 
depleted by more than a factor $\simeq2$.

Finally, we showed that all three measured $^6$Li abundances, as 
well as the upper limits derived, are in excellent agreement
with all standard expectations: standard Big-Bang
nucleosynthesis with $2\leq\eta_{10}\leq6.5$ (Copi et al., 
1995, and references), 'standard' stellar models with respect
to $^7$Li and $^6$Li survival in metal-poor stars, and standard
early galactic evolution of both lithium isotopes (primary 
yields in spallation-fusion processes). The 
graph of $^6$Li abundances {\it vs} $T_{eff}$ obtained is also
in good agreement with the expected shape for $^6$Li isochrones.

\begin{figure}[t]
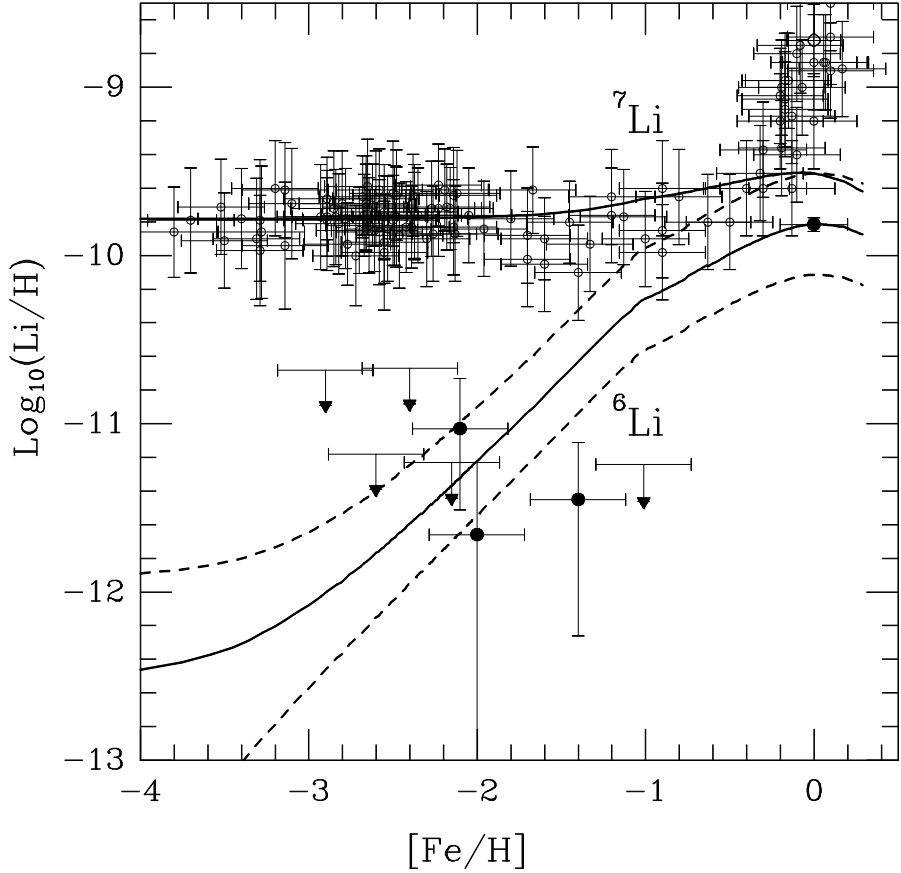

\caption{Graph of $^7$Li and $^6$Li abundances in halo
stars; triangles denote upper limits, filled circles
correspond to actual detections of $^6$Li, and open circles 
to $^7$Li abundances. Error bars are 95\% c.l.,
obtained by summing quadratically statistical and systematics
error bars. Solid curves represent the standard evolutionary
curve for $^6$Li and $^7$Li (no stellar component is included 
for $^7$Li), and the dashed curves delimit the 95\% c.l. on
the $^6$Li prediction.}
\label{Li6_ev}
\end{figure}

\begin{figure}
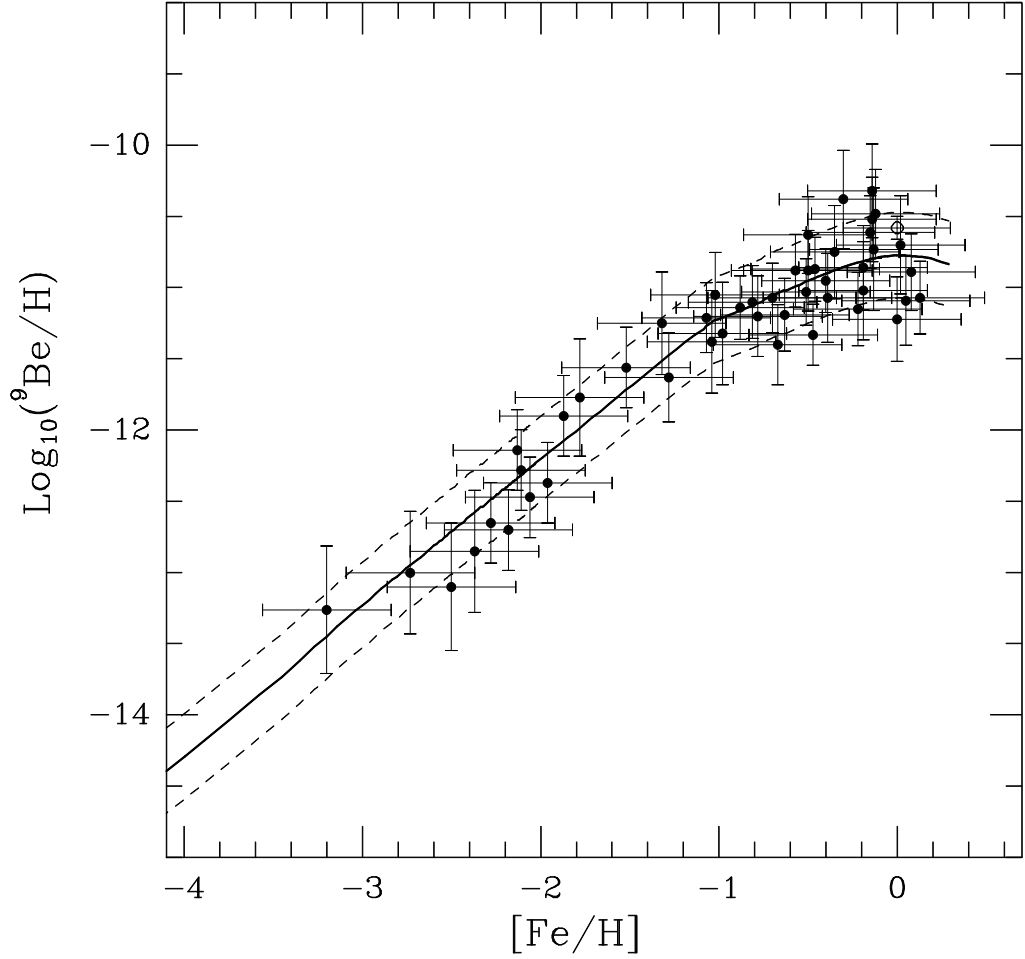

\caption{Graph of $^9$Be abundances {\it vs} metallicity. The 
solid and dashed lines correspond to the $^6$Li 
evolutionary curve and its 95\% c.l. limits (without any primordial 
component) shifted to the
level of $^9$Be abundances.}
\label{Be_ev}
\end{figure}

\begin{figure}
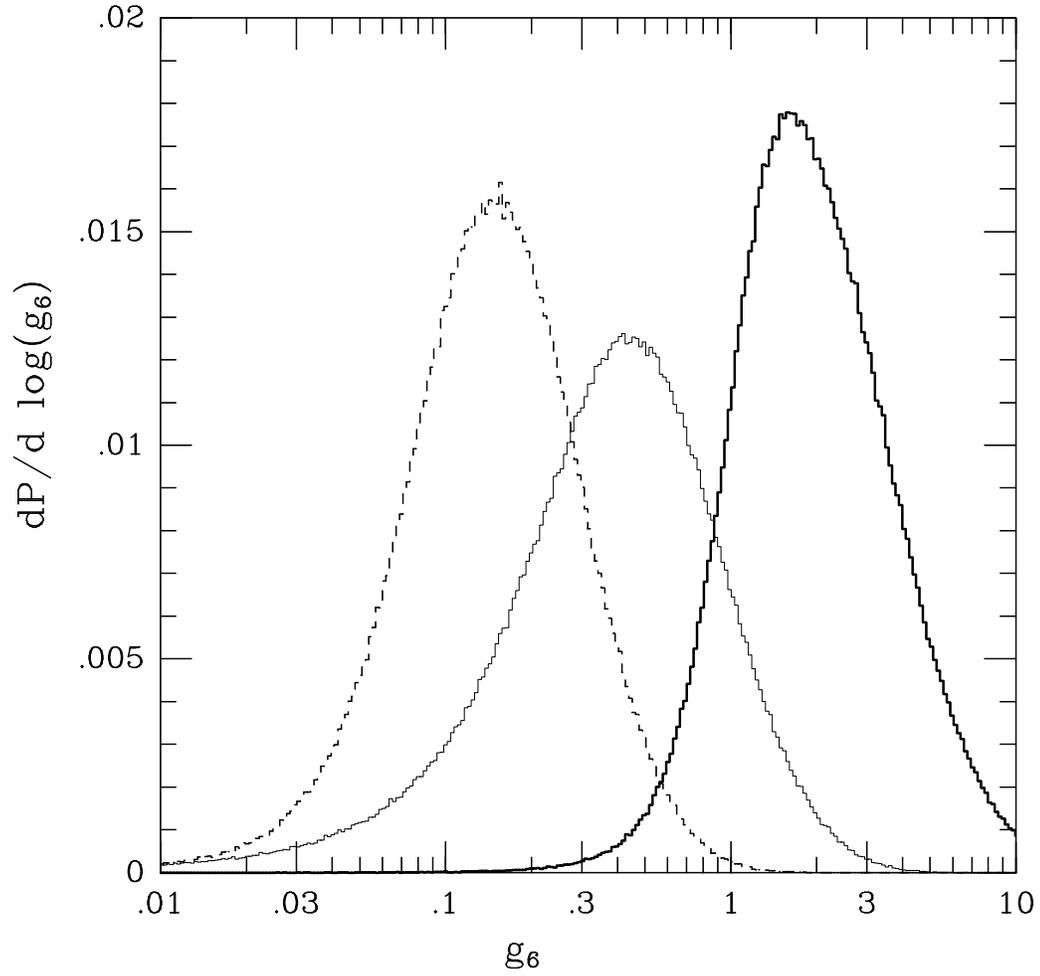

\caption{Histogram of the statistics of the $^6$Li survival
fractions ($g_6$) in HD84937 (heavy solid), HD160617 (light 
solid), and HD201891 (dashed).}
\label{his_g6}
\end{figure}

\begin{figure}
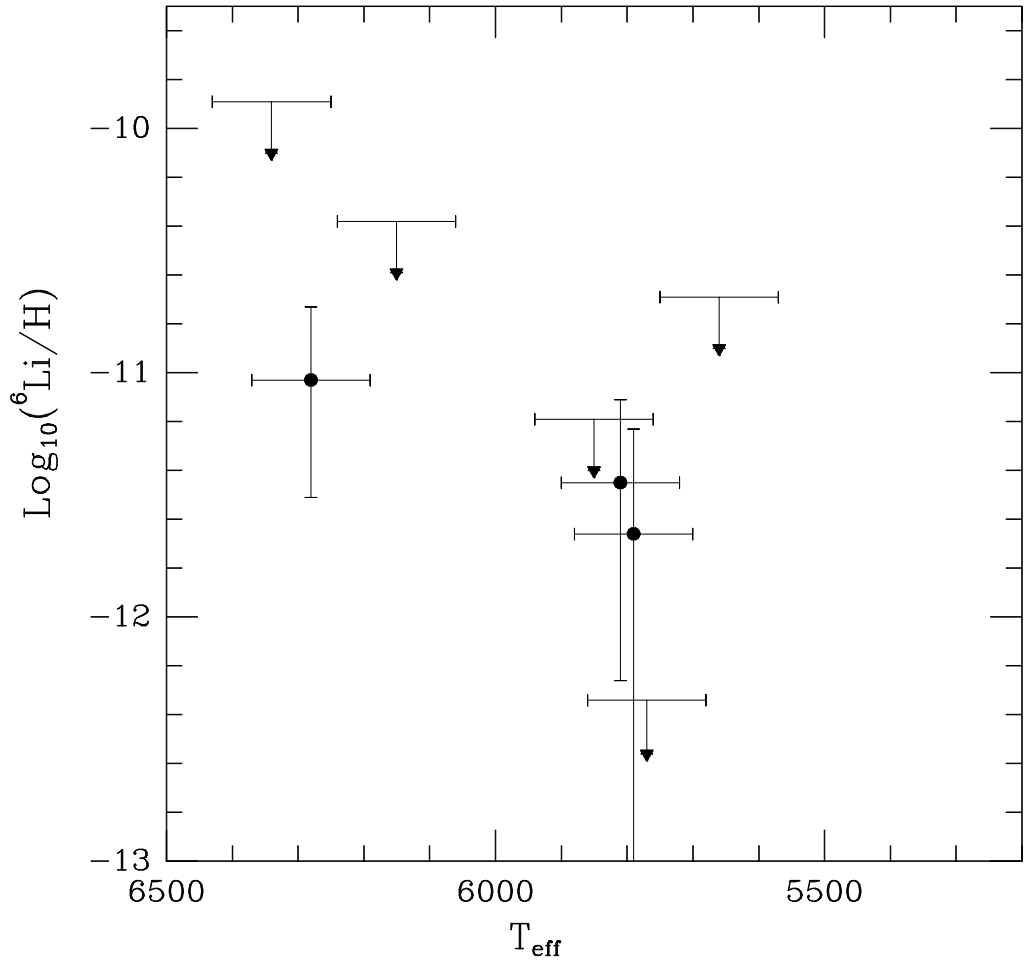

\caption{Plot of $^6$Li abundances {\it vs} $T_{eff}$, after 
having corrected the observed abundances for their trend in 
metallicity, a slope 1 in our case, to the metallicity 
of HD84937, [Fe/H]=$-2.1$. The error bars on 
the abundances implied by this correction were not applied; 
other error bars are 95\% c.l..}
\label{Li6_Teff}
\end{figure}

It is obvious that future measurements of $^6$Li abundances at 
different metallicities would narrow down the uncertainty in
the possible depletion factors for $^6$Li in these stars.
It is equally obvious that such studies would greatly benefit 
stellar modeling physics and the determination of the 
primordial abundance of $^7$Li. We recommend measurements of the 
$^6$Li abundance at a metallicity [Fe/H]$\sim-1$, since this the 
region where $^6$Li detection is favored (see Fig.\ref{Li6_ev}). 
Moreover, such measurements, combined with those at lower 
metallicities, would allow the determination of the slope of the 
correlation Log($^6$Li/H) {\it vs} [Fe/H], which we assumed to be 
$\simeq1$.

We discussed three alternatives to the above scenario.
Such alternatives can only arise from $\alpha-\alpha$ fusion 
creation of $^6$Li, since this is the only channel whereby 
$^6$Li and not $^9$Be is produced. A condition on such 
alternatives is that they have to respect the 
primary behavior of $^9$Be in the early Galaxy.

In the first case, we considered a logarithmic slope for
Log($^6$Li/H) {\it vs} [Fe/H], resulting from an arbitrary
adjustment of the GCR flux to the inverse power of time. We argued
that this scenario lacked any physical basis,
since it would require the early GCR injection flux to be tuned to a 
high power of the supernova rate, in order to reproduce the Be 
and B trends. 

In the second case, where the bulk of LiBeB spallation would 
take place in globular cluster sized objects (Tayler, 1995), no 
firm prediction as to the behavior of $^6$Li can be made at the 
halo-disk transition, hence no evolutionary curve can be
constructed. This scenario remains as a possible loophole for 
obtaining higher depletion factors of $^6$Li in HD84937. We 
note, in passing, that among the models proposed to account for 
the primary behavior of $^9$Be in the halo phase, only this one 
and that of Cass\'e et al. (1995), Vangioni-Flam et al. (1995) 
seem credible at this time. In this latter, C,N, and O nuclei, 
freshly synthesized by supernovae exploding inside their parental 
molecular cloud, are accelerated and eventually decelerate or produce
LiBeB nuclei through spallation on interstellar p and $\alpha$.

Finally, we showed that stellar flare production of $^6$Li 
cannot produce a significant contribution to the observed
$^6$Li, contrary to the suggestion put forward by Deliyannis
\& Malaney (1995).
\bigskip

\noindent{\large\bf Acknowledgments:} we wish to thank C. 
Charbonnel, L. Hobbs, P. Nissen, K. Olive, F. Primas, R. Ramaty,
J. Thorburn, and M. Turner for useful discussions; one 
of us (M.L.) is particularly indebted to M. Cass\'e, F. Spite
and E. Vangioni-Flam for extremely helpful discussions.
This work was supported by the NASA, DoE, and NSF, at the
University of Chicago.

\newpage

\noindent{\Large\bf References}\bigskip

{\leftskip 0.5cm

\bib Anders, E., Grevesse, N.: 1989, Geochim. Cosmochim. Acta
{  53}, 1

\bib Andersen, J., Gustafsson, B., Lambert, D.L.: 1984,
AA {  136}, 65

\bib Boesgaard, A.M., Tripicco, M.J.: 1986, ApJ {  303}, 724

\bib Boesgaard, A. M., King, J.: 1993, AJ {  106}, 2309 

\bib Brown, L., Schramm, D.N.: 1988, ApJ {  329}, L103

\bib Canal, R., Isern, J., Sanahuja, B.: 1975, ApJ {  200}, 646

\bib Canal, R., Isern, J., Sanahuja, B.: 1975, ApJ {  235}, 504

\bib Cass\'e, M., Lehoucq, R., Vangioni-Flam, E.: 1995,
Nature {  373}, 318  

\bib Cecil, F.E., Yan, J., Galovich, C.S.: 1996, preprint

\bib Chaboyer, B.: 1994, ApJ {  432}, L47

\bib Copi, C.J., Schramm, D.N., Turner, M.S.: 1995, Phys. Rev. 
Lett. {  75}, 3981

\bib Deliyannis, C.P., Demarque, P., Kawaler, S.D.: 1990,
ApJS {  73}, 21

\bib Deliyannis, C.P., Malaney, R.A.: 1995, ApJ {  452},

\bib Duncan, D. K., Lambert, D. L., Lemke, M.: 1992, ApJ
{  401}, 584 

\bib Duncan, D.K., Primas, F., Coble, K.A., Rebull, L.M.,
Boesgaard, A.M., Deliyannis, C.P., Hobbs, L.M., King, J., Ryan,
S.G.: 1996a, {\it Cosmic Abundances}, eds. S. Holt, G.
Sonneborn, PASP, in press

\bib Duncan, D.K., Primas, F., Rebull, L.M., Boesgaard, A.M.,
Deliyannis, C.P., Hobbs, L.M., King, J., Ryan, S.G.: 1996b,
in preparation

\bib Epstein, R.I., Arnett, W.D., Schramm, D.N.: 1974, ApJ 190, L13

\bib Feltzing, S., Gustafsson, B.: 1994, ApJ {  423}, 68 

\bib Fields, B. D., Schramm, D. N., Truran, J. W.: 1993, ApJ
{  506}, 559 

\bib Fields, B. D., Olive, K. A., Schramm, D. N.: 1994, ApJ 
{  435}, 185

\bib Fields, B. D., Olive, K. A., Schramm, D. N.: 1995, ApJ 
{  439}, 854

\bib Gilmore, G., Edvardsson, B., Nissen, P. E.: 1992a, AJ
{  378}, 17 

\bib Gilmore, G., Gustaffson, B., Edvardsson, B., Nissen,
P. E.: 1992b, Nature {  357}, 379 

\bib Hobbs, L.M.: 1985, ApJ {  290}, 284

\bib Hobbs L.M., Duncan, D.K.: 1987, ApJ {  317}, 796 

\bib Hobbs, L.M., Thorburn, J.A.: 1994, ApJ {  428}, L25

\bib Lambert, D.L.: 1995, AA {  301}, 478

\bib Lemoine, M., Vangioni-Flam, E., Cass\'e, M.: 1996, ApJ, 
submitted

\bib Mathews, G.J., Schramm, D.N.: 1993,  ApJ, 404, 468

\bib Maurice, E., Spite, F., Spite, M.: 1984, AA {  132}, 278

\bib Meneguzzi, M., Audouze, J., Reeves, H.: 1971, AA
{  15}, 337 

\bib Molaro, P., Bonifacio, P., Castelli, F., Pasquini, L., 
Primas, F.: 1995, {\it The Light Elements 
Abundances}, ed. P. Crane, Springer-Verlag, p.415

\bib Murphy, R.J., Ramaty, R., Kozlovsky, B., Reames, D.V.: 
1991, ApJ {  371}, 793

\bib Nissen, P.E.: 1995, {\it The Light Elements 
Abundances}, ed. P. Crane, Springer-Verlag, p.337

\bib Nissen, P.E., Lambert, D.L., Smith, V.V.: 1994, {\it The 
ESO Messenger}, {  76}, 36

\bib Nollett, K., Schramm, D.N., Lemoine, M.: 1997, in preparation

\bib Olive, K.A., Schramm, D.N.: 1992, Nature {  360}, 439

\bib Pilachowski, C.A., Hobbs, L.M., De Young, D.S.: 1989, ApJ 
{  345}, L39

\bib Pilachowski, C.A., Sneden, C., Booth, J.: 1993, ApJ { 
407}, 699

\bib Pinseonneault, M.H., Deliyannis, C.P., Demarque, P.: 1992, 
ApJS {  78}, 181

\bib Prantzos, N., Cass\'e, M., Vangioni-Flam, E.: 1993, ApJ
{  403}, 630 

\bib Primas, F.: 1995, PhD thesis, Obs. Trieste, Italy

\bib Ramaty, R., Murphy, R.J.: 1987, Space Sc. Rev. {  45}, 213

\bib Ramaty, R., Simnett, : 1991, {\it The Sun in Time}, eds. 
Sonett et al., Arizona, p.232

\bib Ramaty, R., Mandzhavidze, N., Kozlovsky, B., Murphy, R.J.: 
1995, ApJ {  455}, L193

\bib Rebolo, R., Molaro, P., Abia, C., Beckman, J. E.: 1988,
AA {  193}, 193 

\bib   Rebolo, R., Garcia-Lopez, R. J., Perez de Taoro, M. R.:
1995, {\it The Light Elements Abundances}, ed. P. Crane,
Springer, p.420

\bib Reeves, H., Fowler, W. A., Hoyle, F.: 1970, Nature
{  226}, 727 

\bib Reeves, H.: 1994, Rev. Mod. Phys. {  66}, 193 

\bib Ryan, S. G., Norris, J., Bessel, M., Deliyannis, C.:
1992,  ApJ {  388}, 184 

\bib Ryan, S.G., Beers, T.C., Deliyannis, C.P., Thorburn, J.A.: 
1996, ApJ, in press

\bib Smith, V.V., Lambert, D.L., Nissen, P.E.: 1993, ApJ {  
408}, 262

\bib Spite, F., Spite, M.: 1982, AA {  115}, 357 

\bib Spite, M., Fran\c{c}ois, P., Nissen, P.E., Spite, F.:
1996, AA {  307}, 172

\bib Steigman, G., Walker, T. P.: 1992, ApJ {  385}, L13 

\bib Steigman, G., Fields, B. D., Olive, K.A., Schramm, D.N.,
Walker, T.P.: 1993, ApJ {  415}, L35

\bib Tayler, R. J.: 1995, MNRAS {  273}, 215

\bib Thomas, D., Schramm, D. N., Olive, K. A., Fields, B. D.:
1993,  ApJ {  406}, 569 

\bib Thorburn, J.A.: 1994, ApJ {  421}, 318

\bib Vangioni-Flam, E., Cass\'e, M., Audouze, J., Oberto, Y.:
1990, ApJ {  364}, 586

\bib Vangioni-Flam, E., Cass\'e, M., Ramaty, R.: 1995, ApJ, 
submitted

\bib Vauclair, S., Charbonnel, C.: 1995, AA {  295}, 715

\bib Walker, T.P., Matthews, G.J., Viola, V.E.: 1985, ApJ {  
299}, 745

\bib Wheeler, J.C., Sneden, C., Truran, J.W.: 1989, ARAA {  
27}, 279

}
\newpage

{\small
\begin{table}[t]
\caption{Upper limits and measurements of the 
$^6$Li/$^{6+7}$Li ratio in halo stars. Error bars are 95\% c.l. 
estimates; when two error bars are quoted, the first corresponds to
statistical errors, the second to systematics. References: 
{\it (a)} Ryan et al., 1996; 
{\it (b)} Thorburn, 1994; 
{\it (c)} Nissen, 1995; 
{\it (d)} Nissen et al., 1995;
{\it (e)} Smith et al., 1993;
{\it (f)} Hobbs \& Thorburn, 1994;
{\it (g)} Primas, 1995;
{\it (h)} Spite et al., 1995. Evolutionary Status: T-o stands 
for Turn-off, Dw for Dwarf, and Sg for Sub-giant.
}
\begin{center}
\begin{tabular}{|l|c|c|c|c|c|c|}
\hline\hline
Star & $T_{eff}$ & [Fe/H] & N[Li] & ($^6$Li/$^{6+7}$Li) & 
Status & Reference\\
\hline
HD84937 & 6280$\pm150\pm100$ & -2.1$\pm0.16\pm0.2$ & 2.23$\pm0.16\pm0.2$ & 
0.055$\pm0.034$ & T-o &{\it (a),(e),(f)}\\
HD201891 & 5800$\pm150\pm100$ & -1.4$\pm0.16\pm0.2$ & 1.87$\pm0.16\pm0.2$ & 
0.048$\pm0.040$ & Dw &{\it (a),(f)}\\
HD160617 & 5800$\pm150\pm100$ & -2.0$\pm0.16\pm0.2$ & 2.11$\pm0.16\pm0.2$ & 
0.017$\pm0.024$ & Sg &{\it (c),(g),(h)}\\
BD 3$^o$740 & 6340$\pm150\pm100$ & -2.9$\pm0.16\pm0.2$ & 2.21$\pm0.16\pm0.2$ &
$\leq0.08$ & T-o & {\it (a),(f)}\\
BD 26$^o$3578 & 6150$\pm150\pm100$ & -2.4$\pm0.16\pm0.2$ & 2.12$\pm0.16\pm0.2$ &
$\leq0.10$ & T-o & {\it (a),(f)}\\
HD19445 & 5850$\pm150\pm100$ & -2.1$\pm0.16\pm0.2$ & 2.08$\pm0.16\pm0.2$ & 
$\leq0.03$ & Dw & {\it (a),(f)}\\
HD76932 & 5770$\pm150\pm100$ & -1.0$\pm0.16\pm0.2$ & 2.06$\pm0.16\pm0.2$ & 
$\leq0.03$ & Sg & {\it (d),(g)}\\
HD140283 & 5660$\pm150\pm100$ & -2.6$\pm0.16\pm0.2$ & 2.13$\pm0.16\pm0.2$ & 
$\leq0.03$ & Sg & {\it (b),(g),(h)}\\
\hline\hline
\end{tabular}
\end{center}
\label{Li6_data}
\end{table}
}

\end{document}